
\magnification=\magstep1
\overfullrule=0pt\pageno=1
\hsize=15.4truecm
\line{\hfil OCIP/C-92-9}\vskip.5cm
\centerline{\bf QCD CORRECTIONS TO THE LIGHT HIGGS BOSON DECAY INTO}
\centerline{\bf A CHARM- ANTICHARM OR BOTTOM- ANTIBOTTOM PAIR}
\centerline{\bf WITHIN THE MINIMAL SUPERSYMMETRIC MODEL}
\vskip2cm
\centerline{H. K\"ONIG \footnote*{supported by Deutsche
Forschungsgemeinschaft}}
\centerline{Ottawa-Carleton Institute for Physics}
\centerline{Department of Physics, Carleton University}
\centerline{Ottawa, Ontario, Canada K1S 5B6}
\vskip2cm
\centerline{\bf ABSTRACT}\vskip.2cm\indent
We present the results of the contributions
 to the decay rate
of the lightest Higgs particle into a charm-anticharm
and bottom-antibottom pair within the minimal
supersymmetric standard model when
scalar quarks and a gluino are taken within
the relevant loop diagrams.\hfill\break\indent
We show that in the case where the vacuum expectation
values of the Higgs particles obey the condition
$v_1\ll v_2$\ the scalar quarks and gluino change
the decay rate by not more than a few per cent.
\hfill\break\indent
These small contributions make it difficult to
distinguish the Higgs particle of the standard
model from the supersymmetric one.
\vskip2cm
\centerline{ October 1992}
\vfill\break
{\bf I. INTRODUCTION}\hfill\break\vskip.2cm\noindent
It is well known that there are two neutral scalar Higgs
particles $H_1^0,\ H_2^0$\ in the minimal supersymmetric extension
of the standard model (MSSM)[1,2]. $H_1^0$\ has a mass supposed
to be much higher than the Z boson mass $m_{Z^0}$,
whereas the mass of $H_2^0$\ is less than $m_{Z^0}$ at tree
level. Recently it was shown that radiative corrections
coming from the gauge coupling of $H_2^0$\ to the scalar top quark and
its Yukawa coupling to the top quark
enhance its lower mass limit to about 130 GeV [3], which
pushes the possibility of its discovery beyond LEP I but still within
the limits of LEP II.\hfill\break\indent
Because of its relatively small mass $H_2^0$\ is the most likely
Higgs particle to be discovered. The main decay modes of the
light Higgs particle are final state leptons and quarks excluding
the top quark, which we suppose to have a mass around 140 GeV. In
experiment the final state quarks can be distinguished from the
final state leptons. The QCD corrections to the Higgs decay rate
into a charm-anticharm and bottom-antibottom pair
 with a gluon and quarks
within the loop diagram was presented in [4] (and references
therein). There it was shown that these corrections are very
high (several ten per cent).\hfill\break\indent
In this short paper we analyse if the MSSM can lead to any
new measurable contributions to the above decay rate.
The paper is divided into two sections,
in the next section we present and discuss the obtained
results of the caculations whereas in the final section we give
our conclusions and a brief outlook.
\hfill\break\vskip.2cm\noindent
 {\bf II. QCD CORRECTIONS TO $\Gamma(H^0_2\rightarrow q\overline q)$\
WITHIN THE MSSM}\vskip.2cm
In general there are many new diagrams in the MSSM, which
lead to corrections to the decay mode
of the Higgs into a quark-antiquark pair.
There are diagrams with scalar quarks and neutralinos or
charginos within the loop as well as Higgs bosons. All these
diagrams are suppressed by the weak
coupling constant $\alpha_2$. In this paper we are only
interested in the loop diagram with scalar quarks and a gluino,
 which couple to the quarks by the strong
coupling constant $\alpha_s$.
 Although the lightest neutralinos have an
experimental lower mass limit of only around 30 GeV [5], which
is much less than the gluino mass limit of about
70 GeV [6], their contributions are further suppressed by
their diagonalizing angles. But we also consider a very
small gluino mass of 2 GeV still allowed in its low mass
window.\hfill\break\indent
In the Higgs sector of the MSSM there are besides the Higgs mass
parameters, two new angles $\alpha$\ and $\beta$\ coming from
the ratio of the two Higgs vacuum expectation values $v_1$\
and $v_2$ and the mixing angle of the real parts of the
Higgs particles. These two angles are not independent. With
$\tan\beta=v_2/v_1$\ we have the well known relation
$\displaystyle{\tan 2\alpha=\tan 2\beta{{m^2_{H_1^0}+m^2_{H_2^0}}
\over{m^2_{H^0_3}-m^2_Z}}}$\ where $H_3^0$\ is the mass
eigenstate of the imaginary part of the Higgs particles.
In order not to deal with too many parameters we consider
two interesting cases where $v_1\approx v_2$\ (but $v_1$\
smaller than $v_2$) and $v_1\ll v_2$. In the first case
we have $\sin\beta=\cos\beta=\cos\alpha=-\sin\alpha=
1/\sqrt{2}$, whereas in the second case $\sin\beta=1$,
$\cos\beta=0$\ which gives $\alpha=-\pi/2$\ or 0 in special
cases considered in [7]. The consequences to the considered
decay rate are shown below.\hfill\break\indent
In the standard model the Feynman coupling of the Higgs
particle to the fermions is given by $-ig_2m_f/2m_W$\
which in the MSSM has to be changed to
\hfill\break $-ig_2m_f\cos\alpha/
2m_W\sin\beta$\ and  $+ig_2m_f\sin\alpha/
2m_W\cos\beta$\ for the couplings of the light
Higgs particle $H_2^0$\ to the charm-anticharm
and bottom-antibottom pair
respectively.\hfill\break\indent
The diagrams we have to consider for the considered decay mode
are given in Fig.1.
The couplings of $H_2^0$\ to scalar quarks are given in
Fig.110 in [2] and the gluino coupling to scalar quarks
in Eq.C89 in [1].\hfill\break\indent
In our calculations we neglect the charm and bottom quark
masses relative to their scalar partners. This simplifies
the calculations enormously; e.g. we also do not have to
consider mixing terms of the scalar partners of the left
and right handed quarks, which are proportional to the
quark masses -- that is we simply can take $m_{\tilde q_L}
\equiv m_{\tilde q_R}$\ if $\tilde q=\tilde c,\tilde b$\
for the scalar quarks.\hfill\break
The tree level decay rate is given by
$$\Gamma^q_0={{N_CG_Fm_{H^0_2}m_q^2}\over
{4\sqrt{2}\pi}}\beta_0^3\bigl\lbrace {{\cos^2\alpha}
\over{\sin^2\beta}};{{\sin^2\alpha}\over{\cos^2\beta}}
\bigr\rbrace\eqno (1)$$
Here $N_C$\ is the colour factor, $G_F$\ the Fermi constant
and $\beta_0=(1-4m^2_q/m_{H^0_2})^{1/2}$. The first term
in the curly bracket is for the decay rate to charm-anticharm
quarks and the second one for the bottom-antibottom quarks.
The corrections coming from the loop diagrams in the standard
model [4] and those in Fig.1 change $\Gamma^q_0$\ into
$$\Gamma^q=\Gamma^q_0(1+\Delta^{\rm St.}_{\rm rad.}+
\Delta^{\rm SUSY}_{\rm rad.})\eqno (2)$$
$\Delta^{\rm St.}_{\rm rad.}$\ is given in Eq.8 in [4] and
$\Delta^{\rm SUSY}_{\rm rad.}$\ is
$$\eqalignno{\Delta^{\rm SUSY}_{\rm rad.}=&+{8\over 3}
{{\alpha_s}\over{2\pi}}\sin(\alpha+\beta)\bigl\lbrace
{{\sin\beta}\over{\cos\alpha}};{{\cos\beta}\over{\sin\alpha}}
\bigr\rbrace\int\limits^1_0d\alpha_1\int\limits^{1-\alpha_1}_0d\alpha_2
{{-m_Z^2}\over{F_{\tilde g\tilde q}^{H_2^0}}}&(3)\cr
F_{\tilde g\tilde q}^{H_2^0}=&m_{\tilde g}^2-(m^2_{\tilde g}
-m^2_{\tilde q})\alpha_1-(m^2_{\tilde g}-m^2_{\tilde q}
+m^2_{H_2^0}\alpha_1)\alpha_2\cr}$$
The terms in the curly bracket are to be understood as in
Eq.1. The $\alpha_2$- integration can be done easily by
hand and the $\alpha_1$- integration we do numerically.
\hfill\break\indent
First of all we see that the angles $\alpha$\ and $\beta$\
are important. E.g. in the case with $v_1\approx v_2$
($v_1$\ smaller than $v_2$) we have $\sin(\alpha+
\beta)$\ goes to zero and the MSSM does not contribute
at all to the radiative correction of this decay rate.
In the case $v_1\ll v_2$ (that is $\cos\beta$\ goes to
zero) we have $\sin\alpha$\ goes to zero keeping
the charm coupling unchanged
and the bottom coupling at a fixed value.
Because we are only interested in the maximal
influence the MSSM gives to this decay rate
we have to keep $v_1\ll v_2$ and set the values
in the curly brackets in Eq.1 and Eq.3 equal to 1.
\hfill\break\indent
In Eq.9 in [4] the authors kept the strong coupling
constant $\alpha_s$\ as a running function of the
Higgs mass. It is well known that the MSSM changes
the running of the coupling constant. The final value
of $\alpha_s$\ at the weak scale $m_Z$\ depends where
the SUSY breaking scale lies and varies from $\alpha_s=
0.125$\ for the SUSY breaking scale at the
weak scale to $0.118$\ for the SUSY breaking scale at 1 TeV [8]. If we
suppose that the SUSY breaking scale is above 1 TeV the
standard renormalisation equations for the strong
coupling constant apply. To keep our results
independent of the real values of $\alpha_s$\ we
define
$\displaystyle{C^{\rm SUSY}_{\rm max.}\equiv{{\Delta_{\rm rad.}
^{\rm SUSY}}\over{\Delta^{\rm St.}_{\rm rad.}}}}$,
which shows us the maximal radiative correction
of the gluino and scalar quarks in the MSSM compared to the
radiative correction of the quarks and gluon within
the standard model independent of the strong coupling
constant.
\hfill\break\indent
In the following we consider three different cases.
In case I we take the gluino mass and scalar quark
masses to be 80 GeV slightly above the experimental
lower mass limit [6]. In case II we take more
realistic values with $m_{\tilde q}=100$\ GeV and
$m_{\tilde g}=150$\ GeV.
In case III we consider the still experimentally
allowed low mass window of the gluino with
$m_{\tilde g}=2$\ GeV  and $m_{\tilde q}=80$\ GeV,
which leads to the highest contribution.
\hfill\break\indent
In Fig.2 we have plotted $\Gamma_0^c$\ (higher solid line),
$\Gamma^c_{\rm St.}\equiv\Gamma_0^c(1+\Delta^{\rm St.}_{\rm rad.})$\
(lower solid line) and $\Gamma^c$\
in the case I (dashed line), case II (dash-dotted line) and
case III (dotted line). The supersymmetric radiative correction
has a minus sign relative to the standard radiative correction,
which leads to a small enhancement over $\Gamma^c_{\rm St.}$.
The highest contribution is in case III for $m_{H_2^0}=140$\
GeV and lies at 7.45\%.\hfill\break\indent
In Fig.3 we present the same results as in Fig.2 for $\Gamma_0^b$,
$\Gamma^b_{\rm St.}$\ and $\Gamma^b$\ in the three different cases
considered above. Here $\Gamma^b_{\rm St.}$\ is enhanced at most
by 4.99\%.\hfill\break\indent
Finally in Fig.4 we present the absolute values for the $\alpha_s$\
independent parameter $C^{\rm SUSY}_{\rm max}$\ as defined above.
The upper line of identical lines is for the bottom decay mode
and the lower one for the charm one.  $C^{\rm SUSY}_{\rm max}$\
is higher for the bottom quark which leads to a lower enhancement
of $\Gamma^b$\ relative to $\Gamma^c$\ coming from the relative
minus sign of the SUSY correction relative to the standard correction.
\hfill\break\indent
If we take the gluino mass at a very high value (larger than 500 GeV)
we get an enhancement over $\Gamma^{c,b}_{\rm St.}$\ below 1\%.
\hfill\break\indent
As a result even for small masses of the gluino
the decay rate of the Higgs particle
into a charm-anticharm or bottom-antibottom pair turns out to be
a poor experimental quantity to distinguish
the standard Higgs particle from the lightest supersymmetric
one.
\hfill\break\vskip.2cm\noindent
{\bf IV. CONCLUSIONS AND OUTLOOK}\vskip.2cm
Under simplifying assumptions such as neglecting the
charm and bottom quark masses compared to their
supersymmetric partners and in the case $v_1\ll v_2$\
we have shown the maximal contribution of the MSSM
to the decay rate of the lightest supersymmetric
particle into a charm-anticharm or bottom-antibottom
quark pair. We have seen that the contribution of
the scalar quarks and gluino to this decay rate lies
in the range of a few per cent for a light gluino mass
and drops under 1\% if the gluino mass is taken to be heavy
(500 GeV). These small values make the considered decay
rate very unlikely to distinguish the standard
Higgs particle from the lightest supersymmetric one.
\hfill\break\indent
This might be very different if we were considering the decay of
the heaviest neutral scalar Higgs particle $H_1^0$\
into quarks. Here we have to consider many more decay
modes. First of all it might be heavy enough to decay
into a top-antitop quark pair, which makes the
calculation more difficult. Here we cannot neglect the
mixing of the scalar partners of the left and right
handed top quark which becomes proportional to the
top quark mass. Secondly we also have to consider
the decay of $H^0_1$\ into $W^+W^-$-- and $Z^0Z^0$-- bosons.
A calculation of the $H^0_1$\ into $Z^0Z^0$-- bosons was recently
considered in [9]. The authors there have taken quarks
and their supersymmetric partners within the relevant
loop diagrams. However in a complete calculation one
has also to take charginos and neutralinos as well as
Higgs bosons into account. These particles might give
significant contribution to this
decay mode and should also be considered [10].
\hfill\break\vskip.2cm\noindent
{\bf IV. ACKNOWLEDGEMENTS}\vskip.2cm
The author would like to thank P. Kalyniak for
useful discussions. This work was supported by
the Deutsche Forschungsgemeinschaft.\hfill\break
\vfill\break
{\bf REFERENCES}\vskip.2cm
\item{[\ 1]}H.E. Haber and G.L. Kane, Phys.Rep.{\bf 117}(1985)75.
\item{[\ 2]}J.F. Gunion et al,{\it "The Higgs Hunter's Guide"},
Addison-Wesley Publishing Company, Redwood City, CA 1990.
\item{[\ 3]}H. Haber and R. Hempfling, Phys.Rev.Lett.
{\bf 66}(1991)1815;\hfill\break
Y. Okada et al, Prog.Theor.Phys.{\bf 85}(1991)1; Phys.Lett.
{\bf B262}(1991)54;\hfill\break
J. Ellis et al, Phys.Lett.{\bf B257}(1991)83;\hfill\break
for a more general analysis leading to a higher mass limit of
about 150 GeV see G.L. Kane et al, "Calculable upper limit
on the mass of the lightest Higgs boson in any perturbativly
valid supersymmetric theory", October 1992,
Bulletin-board: hep-ph@xxx.lanl.gov- 9210242.
\item{[\ 4]}P. Kalyniak et al., Phys.Rev.{\bf D43}(1991)3664.
\item{[\ 5]}G. Wormser, "Searches for SUSY particles at LEP",
October 1992, LAL-92-56.
\item{[\ 6]}Review of Particle Properties,
 Phys.Rev.{\bf D45}(1992)Part II.
\item{[\ 7]}J.F. Gunion and H.E. Haber, Nucl.Phys..{\bf B278}
(1986)449.
\item{[\ 8]} P. Langacker and M. Polansky, "Uncertainties
in Coupling Constant Unification", October 1992, UPR-0513T,
Bulletin-board: hep-ph@xxx.lanl.gov- 9210235;
 P. Langacker, "Proton
 Decay", Bulletin-board: hep-ph@xxx.lanl.gov- 9210238;\hfill\break
V. Barger et al., "Supersymmetric Grand Unification: Two Loop
Evolution of Gauge and Yukawa Couplings", September 1992,
Mad/PH/711, Bulletin-board: hep-ph@xxx.lanl.gov- 9209232.
\item{[\ 9]} D. Pierce and A. Papadopoulos, "Radiative
Corrections to the Higgs boson decay $\Gamma(H\rightarrow ZZ)$\
in the minimal supersymmetric model", June 1992, UCB-PTH-92/23,
LBL-32498, Bulletin-board: hep-ph@xxx.lanl.gov- 9206257.
\item{[10]} H. K\"onig, "Higgs boson decay into
$t\overline t$ quarks,
$W^+W^-$ -- and $Z^0Z^0$-- bosons within the minimal supersymmetric
standard model", in preparation.
\vfill\break
{\bf FIGURE CAPTIONS}\vskip.2cm
\item{Fig.1}The diagrams with scalar quarks and gluino
within the loop, which contribute to the decay
rate of the lightest Higgs particle into a
 quark-antiquark pair within the MSSM.
\item{Fig.2}$\Gamma^c_0$\ (upper solid line), $\Gamma^c_{\rm St.}$\
(lower solid line) and $\Gamma^c$\ as a function of
$m_{H_2^0}$. The $\Gamma^c$\ are: Case I (dashed line),
case II (dash-dotted line),
case III (dotted line).
\item{Fig.3}As in Fig. 2 for the b-quark.
\item{Fig.4} The absolute value of
$\displaystyle{C^{\rm SUSY}_{\rm max.}}$\ as a function of
$m_{H^0_2}$. Lines for the various cases are
as in Fig.2. The upper line of the same line is
for the b-quark and the lower one for the c-quark.
\vfill\break
\end